\documentclass[twocolumn,secnumarabic,amssymb,nobibnotes,showpacs,aps,prl]{revtex4}

\usepackage{amsmath,amssymb}
\usepackage{array}
\usepackage{graphicx}
\usepackage{color}
\usepackage{units}
\usepackage{natbib}

\newcommand{\eref}[1]{(\ref{#1})}

\newcommand{\rem}[1]{}

\newcommand{\coupling} {\varepsilon}

\newcommand{\figsetup}{1}
\newcommand{\figphasesynchronization}{2}
\newcommand{\figarnold}{3}
\newcommand{\figreconstruction}{4}

\begin{document}

\title{Synchronization of  Sound  Sources}%

\author{Markus Abel}
\author{Karsten Ahnert}
\author{Steffen Bergweiler}
\email[corresponding author: ]{markus.abel@physik.uni-potsdam.de}
\affiliation{Department of Physics and Astronomy, Potsdam University, Karl-Liebknecht-Str. 24, D-14476, Potsdam-Golm, Germany}
\date{\today}%

\begin{abstract}

Sound generation and -interaction is highly complex, nonlinear and
self-organized. Already 150 years ago Lord Rayleigh raised the following
problem: Two nearby organ pipes of different fundamental frequencies sound
together almost inaudibly with identical pitch. This effect is now understood
qualitatively by modern synchronization theory
(M. Abel et al., J. Acoust. Soc. Am., 119(4), 2006).
For a detailed, quantitative investigation, we substituted one pipe by an
electric speaker.  We observe that even minute driving signals force the pipe
to synchronization, thus yielding three decades of synchronization -- the
largest range ever measured to our knowledge.  Furthermore, a mutual silencing
of the pipe is found, which can be explained by self-organized oscillations,
of use for novel methods of noise abatement. Finally, we develop a specific
nonlinear reconstruction method which yields a perfect quantitative match of
experiment and theory.
\end{abstract}

\pacs{43.25.+y, 05.45.Xt, 07.05.Kf, 07.05.Tp}
\maketitle

%

\maketitle

\paragraph*{Introduction.}

In a seminal publication Lord Rayleigh reported on experiments involving two
organ pipes of close pitch, positioned next to each other. He observed the
following peculiar behavior: alone, each pipe sounded with its own natural
frequency. Together they sounded in perfect unison and almost reduced one
another to silence \cite{Rayleigh-82}. This phenomenon can be described by
synchronization theory, a general nonlinear principle with striking
applications in the natural sciences including physics, chemistry, neurology or
biology \cite{Pikovsky-Rosenblum-Kurths-01}.

Here, we focus on an organ pipe as i) prototypical for aero-acoustical sound
generation, ii) paradigmatic for synchronization of coupled oscillators and
iii) a beautiful musical instrument for which we present a mathematical
model. By driving one organ pipe sinusoidally by an electric speaker we can
demonstrate in great detail the importance of nonlinear effects in sound
generation and -interaction \cite{Fletcher-NL} in contrast to linear response
theory \cite{RossingFletcher-97} conventionally applied for musical
instruments.

The general description of the dynamics of an organ pipe is given by the
compressible Navier-Stokes Equations with suitable boundary conditions. One
can solve the equations numerically \cite{CAA}, or investigate them
analytically \cite{Howe-03}. Both ways reproduce different aspects of sound
production.  Here, we are interested in the {\it interaction} of an organ pipe
with a sound source. Then, it is of advantage to model only the relevant
characteristics in terms of reduced models \cite{Fabre-00}. Such an elementary
model allowing for complex dynamics is given by an autonomous oscillator
\cite{Pikovsky-Rosenblum-Kurths-01,Nayfeh-Mook-79}, which includes an
oscillatory unit, energy supply, and energy loss by radiation and damping.

Let us identify these units in the organ pipe. Energy is supplied steadily by
the wind system through the pipe foot and establishes a turbulent vortex
street. Each time a vortex detaches, a pressure fluctuation enters the
resonator, inside which characteristic waves are selected, and radiated at the
pipe mouth by an oscillating air-sheet \cite{Fabre-00,Fabre-Hirschberg-96}. In
our model, this air-sheet constitutes the basic oscillating unit. Inside the
resonator Sound Pressure Levels (SPLs) up to \unit[160]{dB} can occur, such
that viscous damping contributes in energy dissipation.  External acoustical
fields couple to the system through the air-sheet, possibly described by a
(nonlinear) acoustical admittance \cite{Thwaites-Fletcher-83}. By Lighthills
analogy \cite{Howe-03}, a coupling by the turbulent vortex street is expected
to be of lower order.

The above scenario can be  described by a reduced, two-dimensional
model for the oscillatory unit $\xi$:
\begin{equation}
    \ddot{\xi} - g(\xi,\dot{\xi}) = 0
    \label{eq:Osc}
        \text{,}
\end{equation}
where the function $g(\xi,\dot{\xi})$ contains the above mentioned ingredients
and with the condition that a limit-cycle solution of frequency $\nu_0$
exists. Then, with $\xi=A(t)\cdot e^{i\phi(t)}$ the phase $\phi$ and the
amplitude $A$ are well defined. An external driving enters on the right hand
side of Eq.~\eref{eq:Osc}. Corresponding to the experiment, sinusoidal driving
is used: $\coupling \sin(2\pi \nu t + \phi_0)$ with $\coupling$ the coupling
strength. Here, one assumes that an oscillator represents the basic physics of
the pipe with regard to sound generation and synchronization - typically, the
oscillating air sheet \cite{Fabre-00}.  Because the air sheet is the source of
sound radiation, the measurement at the microphone can be taken as the state
of the oscillator (with a phase shift accounting for distance).

Close to the limit cycle the amplitude is slaved by the phase, allowing the
description in terms of the phase difference, $\varPsi$, between driving and
oscillator
\begin{equation}%
\dot{\varPsi} = - 2\pi (\nu -\nu_0) + \coupling\, q(\varPsi)
\text{.}
\end{equation}
The parameters are driving frequency $\nu$, and coupling $\coupling$.
The study of the parameter plane $(\nu,\coupling)$ yields triangular-shaped
synchronization regions, the  well--known Arnold-Tongues \cite{Arnold-91}.

In order to determine Eq.~\eref{eq:Osc} {\it directly} from data, we have
elaborated a numerical method based on embedding theory
\cite{Sauer-Yorke-Casdagli-91} which allows for a reconstruction and
comparison of the characteristics of model and data. We recover the power
spectra and synchronization properties of the organ as an acoustical
systems. This opens the way for a detailed theoretical investigation of the
system and hints to how a model can be derived from first principles.

\paragraph*{The experiment.}

\begin{figure}
\begin{center}
\includegraphics[draft=false,width=0.35\textwidth]{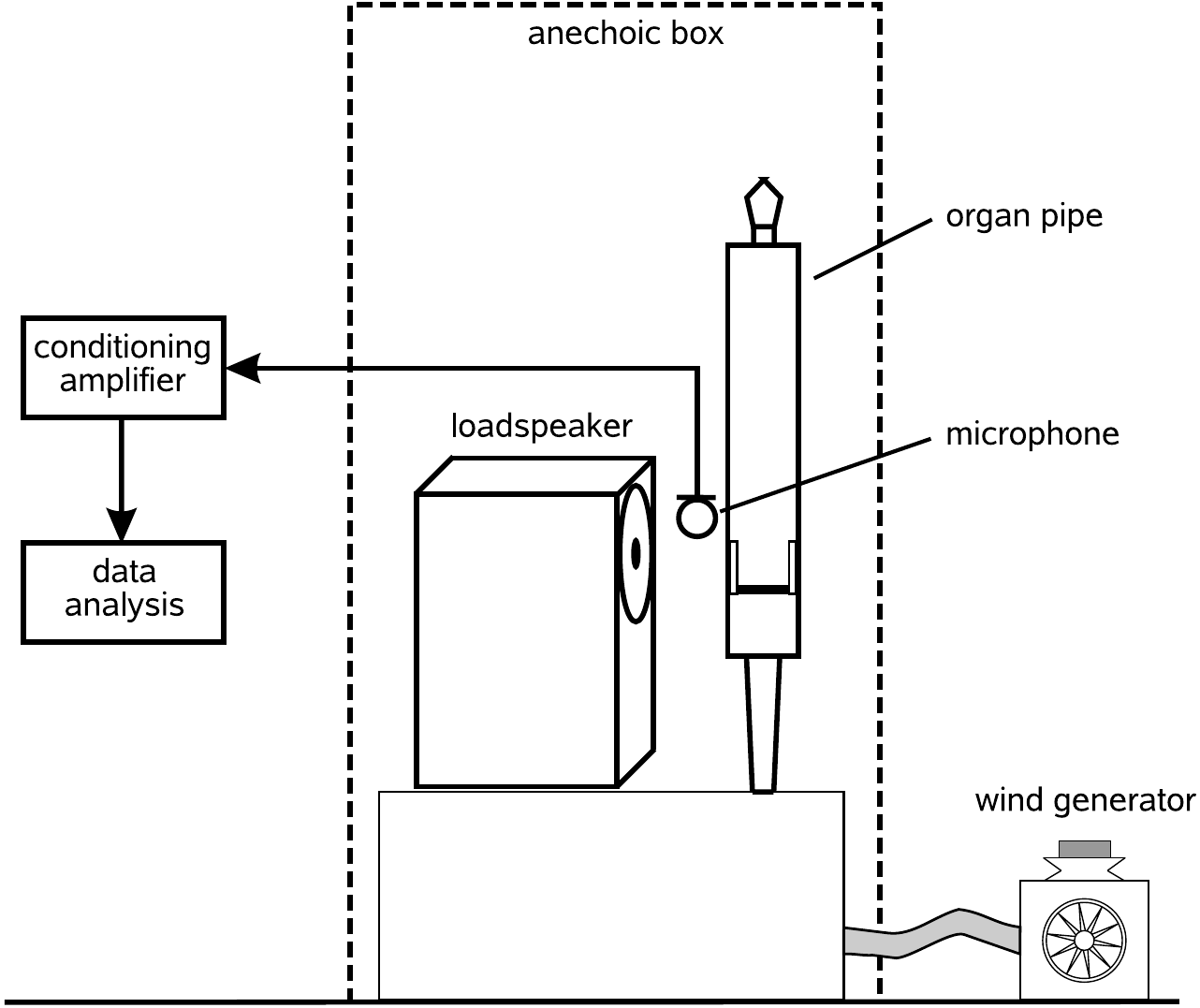}
  \caption{Sketch of the experiment. Pipe and loudspeaker stand side by side,
    the signal is measured by a microphone with equal distance to both sound
    sources.}
\end{center}
\end{figure}

The setup of the experiment is sketched in Fig.~\figsetup. The pipe
\cite{Schuke} was wooden and closed at the upper end, tuned at
$\nu_0=686\text{Hz}$. It was driven by an especially fabricated miniature
organ \cite{Schuke} with a blower connected to the wind-belt and further by
flexible tubes to the wind-chest. Measurements took place inside a suitable
anechoic box. The loudspeaker for the external, driving sound signal, was
positioned side-by-side to the organ-pipe, cf. Fig.~\figsetup. 
The emitted signals were registered at a
distance of \unit[16]{mm} to either pipe and speaker, allowing for
constructive or destructive interference of the superposed signals. To ensure
that the phase of the pipe is correctly detected, we carried out simultaneous
control measurements inside and outside the pipe and at the microphone -- all
results were consistent.

To explore the coupling-detuning plane $(\nu,\coupling)$, the loudspeaker $SPL$
and frequency were varied separately; the first between \unit[+10]{dB} and
\unit[-50]{dB} relative to the reference signal of the organ pipe in steps of
\unit[2]{dB}, the latter according to the size of the synchronization
range. To determine the minimal achievable resolution in $\nu$, we consider
the sources of variations of the pipe's frequency. The wind pressure was
\unit[700$\pm$9]{Pa}, giving a frequency variation of \unit[$\pm$0.1]{Hz} at
the scale of seconds; the temperature varied with the circadian rhythm at
\unit[292$\pm$1]{K} with a resulting variance of $\simeq$\unit[2]{Hz}
\cite{Bohn-88}. This slow change did not affect an individual run with fixed
$SPL$, however, runs with different $SPL$ were adjusted to the circadian
variation.  We measured the synchronization range down to the maximal
resolution of \unit[0.1]{Hz} as set by the wind supply; below noise destroys
synchronization. With respect to noise, we enhanced the signal-to-noise ratio
was enhanced by long-time averaging, such that irregular phase slips are
leveled out and a synchronization region can be obtained even for very small
driving. This was achievable by heavily automatized measurements of a total
duration of about 3 weeks.

\begin{figure}
\begin{center}
\includegraphics[draft=false,width=0.35\textwidth]{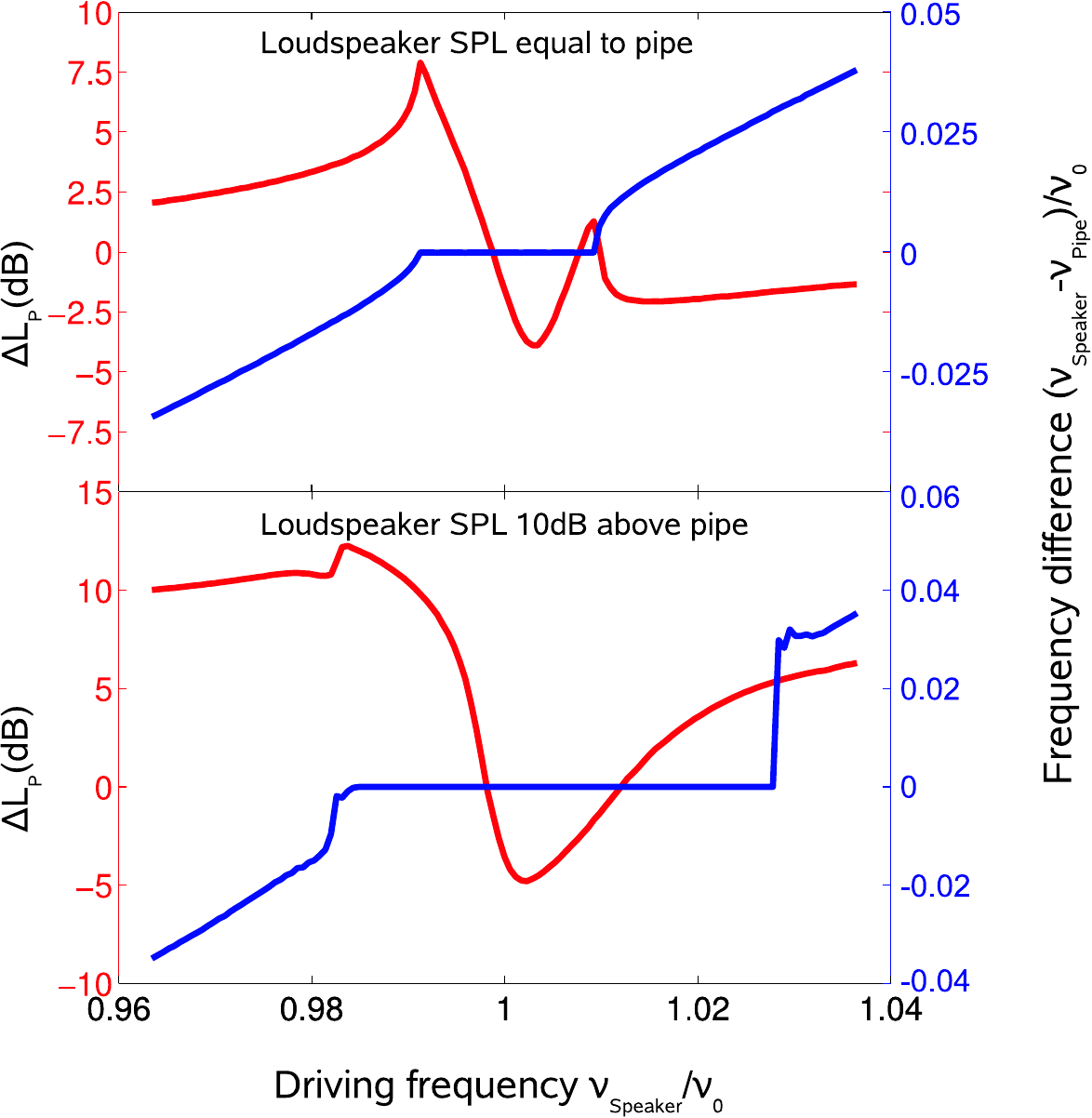}
  \caption{Synchronization plot: if the detuning of pipe and loudspeaker is
    small enough, frequency synchronization is found as a plateau.  Within the
    synchronization region, the phase difference $\varPsi$ varies over an
    interval of size $\pi$, where the measured $SPL$ shows a sharp minimum for
    $\varPsi=\pi$.  Top ($SPL_{Speaker}=SPL_{Pipe}$): the frequency shows a
    very nice plateau with a saddle-node bifurcation; for $\varPsi=\pi$
    negative interference is observed by a decrease of \unit[6]{dB}.  Bottom
    ($SPL_{Speaker} \simeq SPL_{Pipe} \unit[+10]{dB}$): the transition from
    synchronization is quite sharp, indicating strong
    nonlinearities. The amplitude decrease can be explained by synchronization
    with an additional Helmholtz-resonator-like behavior of the pipe,
    deepening the $SPL$ gap to \unit[15]{dB}. The maximum is obtained by
    addition of the amplitudes to \unit[12]{dB}.}
\end{center}
\end{figure}

We investigated two acoustically relevant characteristics: the
frequency difference,
$\Delta\nu/\nu_0=(\nu_{Speaker}-\nu_{Pipe})/\nu_0$, and the spectrum of the measured loudspeaker signal.
Their dependence on the detuning is shown in
Fig.~\figphasesynchronization\ for two exemplary relative driving strengths,
\unit[0]{dB} and \unit[10]{dB}. Note, that $\nu_{Pipe}$ is the frequency of
the {\it driven} pipe which might be different from its natural frequency.

The transition to synchronization can be seen from the graph for the frequency
difference. For equal $SPL$ and below (upper panel in
Fig.~\figphasesynchronization), a saddle node bifurcation is found, as
predicted theoretically
\cite{Abel-Bergweiler-Multhaupt-06,Pikovsky-Rosenblum-Kurths-01}.  For
couplings stronger than \unit[0]{dB}, the bifurcation tends to become very
sharp, indicating that the weakly nonlinear approximation
\cite{Pikovsky-Rosenblum-Kurths-01} breaks down (lower panel
Fig.~\figphasesynchronization).

In the graphs for the amplitude, one recognizes two effects: synchronization
and resonance. Due to synchronization the phase shift between the two emitted
signals varies within an interval of size $\pi$; as for Helmholtz-resonance,
the loudspeaker signal is phase-shifted by $\sim\pi$ and re-emitted by the
pipe. From synchronization alone, the superposition of the signals of speaker and pipe can only vary within
$SPL_ {Speaker}\pm SPL_{Pipe}$; with resonance much weaker amplitudes result,
as seen in Fig.~\figphasesynchronization. This result implies a novel way of
sound reduction, where the reductor acts as an active element, adjusting the
frequency without any external control.

\begin{figure}
  \includegraphics[draft=false,width=0.45\textwidth]{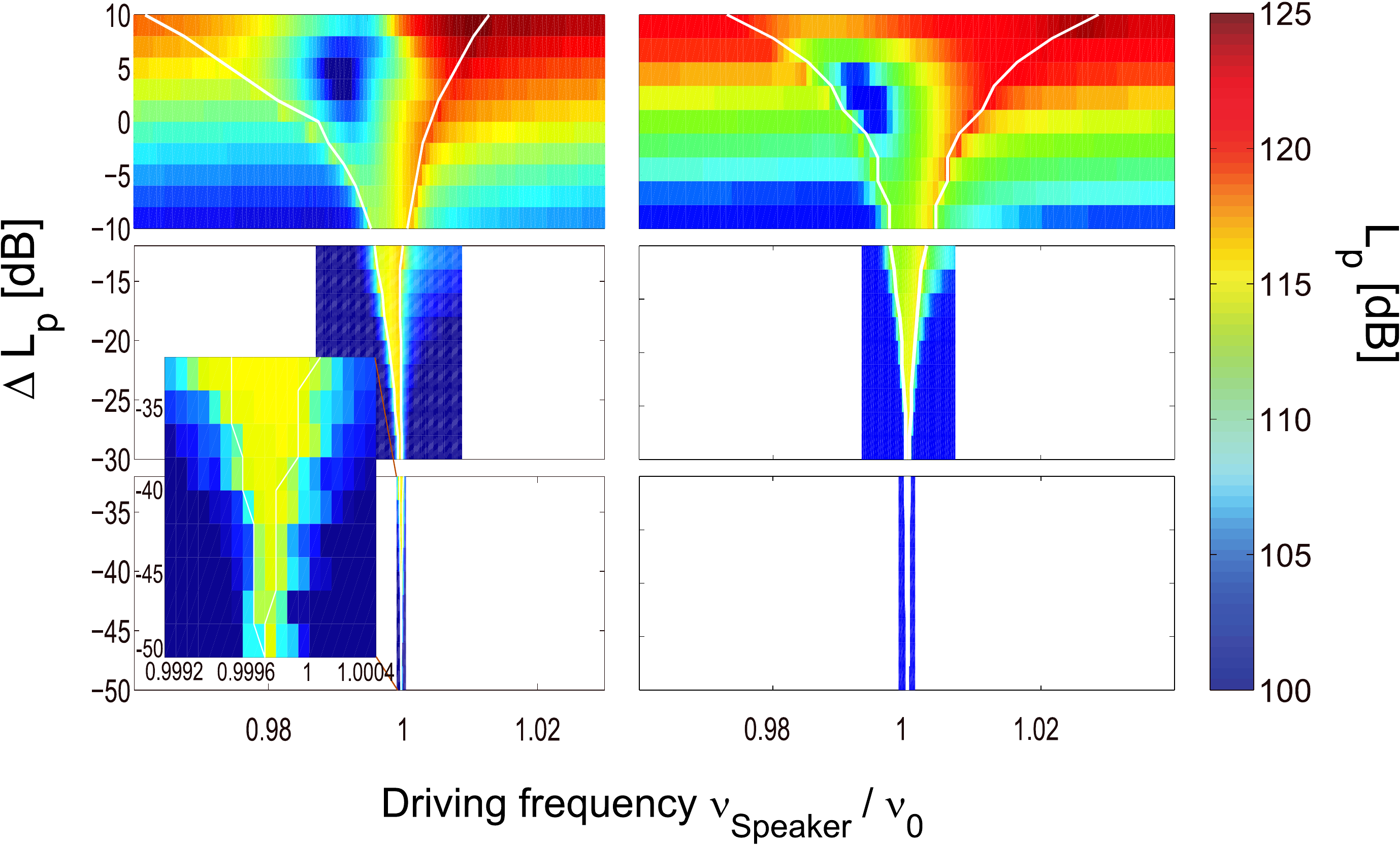}
  \caption{Arnold tongue measured over approximately 3 decades. The left plot
    shows the Tongue obtained from the experimental data, while the right one
    shows the reconstructed tongue. The color coding corresponds to the
    amplitude measured at the microphone position at the loudspeaker
    frequency $S(\nu_{Speaker})$. To guide the eye, the synchronization region is marked by a
    straight line. The synchronization edge grows linearly with the coupling,
    seen as a logarithmic bend in the used semi-log plot.  The lowest possible
    frequency resolution, \unit[0.1]{Hz}, corresponds to the variation of the
    frequency produced by the wind supply, shown by the inset. The
    reconstructed Arnold tongue is much more symmetric. This indicates again
    strong nonlinearities. 
	The coincidence between
    experiment and model is almost perfect in the low-coupling region.}
\end{figure}

The full variation in parameter space is shown in the left panel of Fig.~\figarnold\ by an
Arnold Tongue, here plotted logarithmically due to the enormous coupling range
investigated. The color coding corresponds to the the spectral power of the
loudspeaker frequency $S(\nu_{Speaker})$, measured at the microphone.  We
varied the amplitude of the loudspeaker from \unit[+10]{dB} to \unit[-50]{dB},
relative to the $SPL$ of the organ pipe, in steps of \unit[2]{dB}. The range
covers 3 decades -- the widest range ever measured in synchronization
experiments. The synchronization edges are marked in Fig.~\figarnold\ by white
lines. Clearly, the linear shape is bent due to the semi-log plotting.

\paragraph*{Reconstruction of the dynamical system.}

We do not have access to the ``state'' of the oscillating air--sheet,
but we know the recorded $SPL$, $x(t)$, of the organ pipe. By embedding theory 
we can infer a differential embedding
$(x,\dot{x},\ddot{x},\dots)$ with a maximal embedding dimension of three
\cite{Sauer-Yorke-Casdagli-91,Tufillaro-Wyckoff-Brown-Schreiber-Molteno-95}.
In this space there exists an equivalent to Eq.~\eref{eq:Osc}:
$\ddot{x}-f(x,\dot{x})=0$. We reconstruct this equation step by step to
further investigate its dynamical and predictive properties.

The data series consists of \unit[110250]{} data points, with a sampling rate
of \unit[11025]{Hz}. Normalization in space and time yields variance $0.5$ and
frequency $1$.
The crucial computation of derivatives was accomplished by spectral smoothing
\cite{Ahnert-Abel-07}: {\it i}) Fourier transformation, {\it ii}) 8th order
Butterworth--filtering with cutoff at \unit[4.5]{$\nu_0$} to suppress noise
amplification , and {\it iii}) back-transformation. If the cutoff is increased
more harmonics enter the filtered time series, if the cutoff is too low too
much information is filtered out, i.e. the necessary nonlinearities are
suppressed.  Thorough testing yielded that in our case a three dimensional
embedding do not improve the results.

The unknown function $f$ is estimated by nonparametric regression, formulated
as a minimization problem: $\|
\ddot{x}-f(x,\dot{x})\|_2\stackrel{!}{=}\text{min}$, with $\|\cdot\|_2$ the
$l^2$-norm of the data vector. The unknown function $f$ is found by variation
in function space, where we used, for the sake of computational simplicity,
polynomials of order three
\cite{Abel-Ahnert-Mandelj-Kurths-05,Gouesbet-Letellier-94}; higher orders do
not improve the model. Specifically,
$f(x,\dot{x}) = a_0 +
a_1 x +
a_2 x^2 +
a_3 x^3 +
a_4 \dot{x} +
a_5 \dot{x}^2 +
a_6 \dot{x}^3 +
a_7 x \dot{x} +
a_8 x^2 \dot{x} +
a_9 x \dot{x}^2
$
with
$
a_0=0.25,
a_1=-0.92, 
a_2=-0.18, a_3  =  -0.12, 
a_4=0.20,
a_5=-0.33, 
a_6=0.056,
a_7=-0.015, 
a_8=-0.923, 
a_9=-0.072$
. Note the striking similarity to the van der Pol oscillator
with $f_{vdP}=-x+\dot{x}(1-x^2)$, reflected in the dominant coefficients $a_1$ and
$a_8$. The latter is responsible for nonlinear damping, whereas energy 
is supplied by the constant and the $\dot{x}$-term. Other terms assist
nonlinear damping and are indispensable to find the correct frequencies in the
reconstructed system. Since the observables are not directly related to the
physical driving (wind) and damping mechanisms we hesitate giving a complete
physical interpretation.

Numerical stability analysis yields a repelling fixed point at $(0.254,0)$ and
an attracting limit cycle, plotted in the inset of Fig.~\figreconstruction,
which shows convincing coincidence with the filtered experimental data. For
acoustical comparison, we compare the power spectra of pipe and reconstruction
in Fig.~\figreconstruction. The positions of the harmonics are in perfect
agreement, and their ratio coincides well.

\begin{figure}
  \begin{center}
    \includegraphics[draft=false,width=0.25\textwidth]{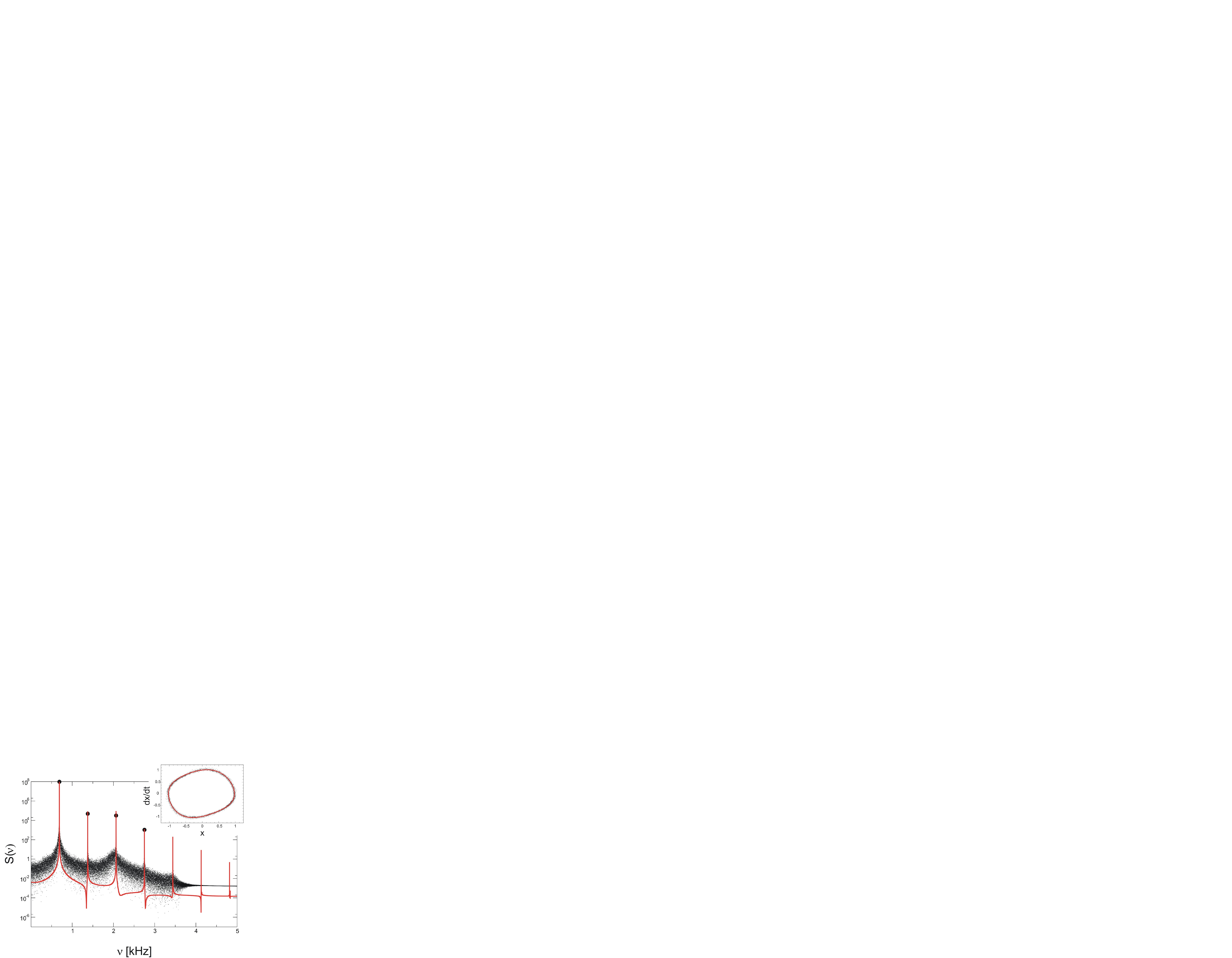}
    \includegraphics[draft=false,width=0.22\textwidth]{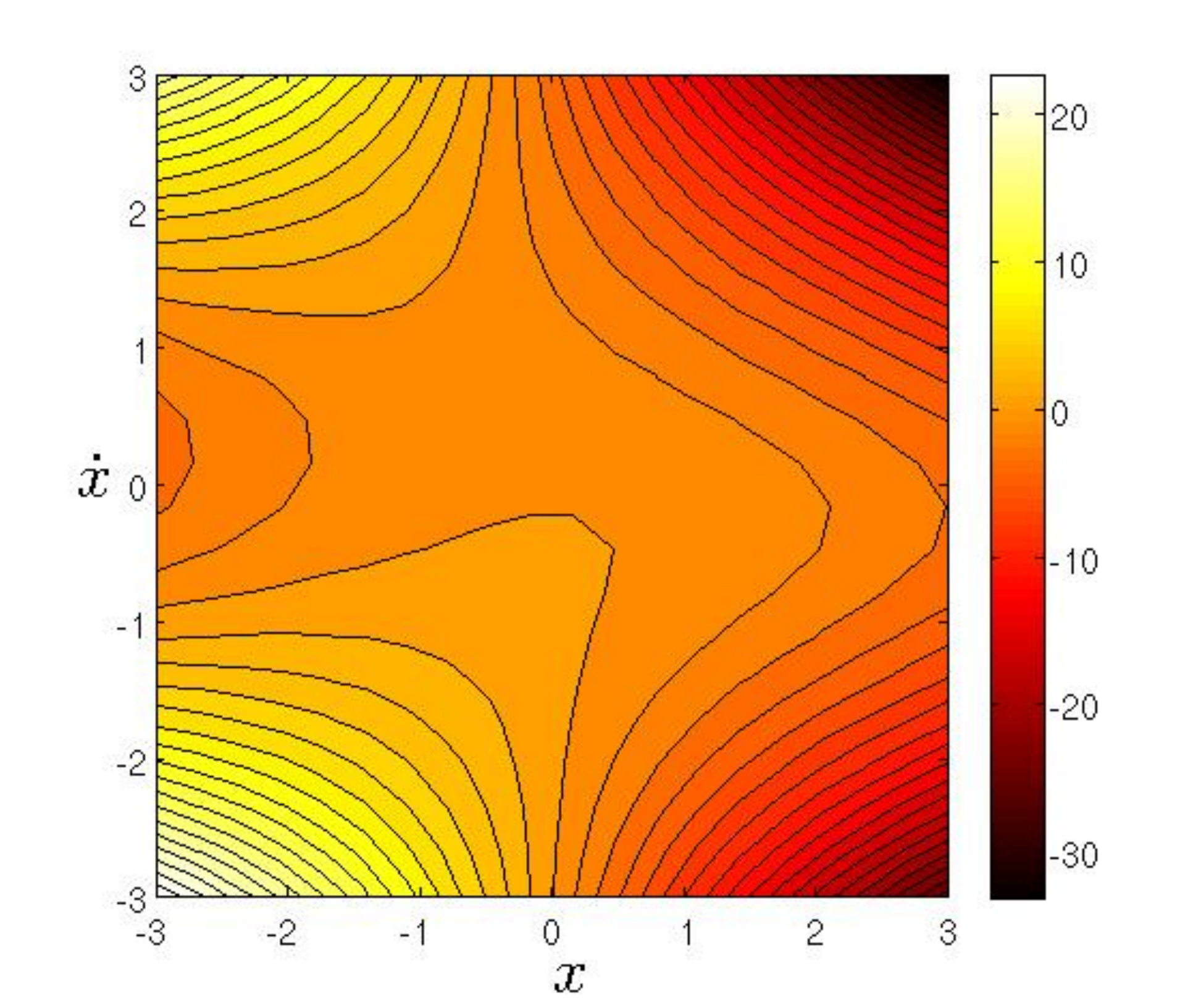}
  \end{center}
  \caption{Left: The power spectra of the measured (black) and the
    reconstructed signal (blue). The musical sound is reproduced quite nicely,
    as recognized by the coincidence of the spectra; for better visibility the
    maxima of the measured signal are shown by black dots. The inset shows the
    time series and its reconstruction in the embedding space, which coincide
    very well. Right: Contour plot of the  function $f$ found by nonparametric regression, the color coding is shown by the colorbar on the right.}
  \label{fig:func}
\end{figure}

Finally, the model is synchronized with an external, sinusoidal driving. For
the coupled equation, we solve  $\ddot{x} = f(x,\dot{x}) +
\epsilon \sin(\omega t)$, with the driving frequency $\omega$ and the coupling
parameter
$\epsilon = 0.025 \cdot SPL_{Speaker}$.
The latter relation describes the influence of the loudspeaker on the pipe
with a factor 0.025, determined in order to obtain optimal coincidence of
experimental and reconstructed Arnold Tongue, see Fig.~\figarnold. Recently, a
way to extract the coupling from data has been proposed
\cite{Kralemann-Cimponeriu-Rosenblum-Pikovsky-Mrowka-08}; in principle it can
be obtained by a detailed analysis of acoustics and fluid dynamics at the pipe
mouth.

\paragraph*{Conclusion.}

Since the time of Lord Rayleigh the nonlinear interaction of acoustical
sources is under discussion \cite{Rayleigh-82}. We highlight the acoustical
effects of phase synchronization, experimentally realized by an organ pipe
driven externally by a loudspeaker. The application of synchronization theory
suggests a novel type of sound control, where the passive element adjusts its
frequency {\it exactly} to the source.  Since the system involved can be
abstracted, we conclude that such a control can be applied to a diversity of
situations, from musical instruments to noise reduction in vibrating systems,
be it mechanical or hydromechanical.

With respect to synchronization, we found the deepest Arnold tongue ever seen
experimentally, suggesting wind instruments as paradigmatic for
synchronization.  The results confirm the theory for small coupling; for
large coupling we give experimental access to nonlinear correction terms to be
analyzed further. To analyze the acoustical properties of the pipe we propose
an autonomous oscillator, reconstructed from a novel type of data
analysis. The agreement between model and experiment in terms of Arnold
Tongues and power spectra is excellent -- musical and synchronization
characteristics are well reproduced. Conventional methods, as transfer
functions, or admittance \cite{Cox-DAntonio-04} do not allow such a direct
interpretation

In this work acoustics is paired with nonlinear dynamics and data
mining.  Organ builders have developed complicated empirical rules to
arrange organ pipes within a register to avoid synchronization
effects. Our results do not only allow for an easy simulation of
instruments, but as well development cycles and tuning of instruments
could be enhanced. Noise reduction is possible by self-organization of
two sound sources such that they interfere negatively - no external
control is needed.  The applications of this principle might be
interesting for a variety of technically important situations.

\paragraph*{Acknowledgments.}
We acknowledge inspiring discussion with A. Pikovsky and M. Rosenblum, and thank F. Spahn for many suggestions. The organ builder company Schuke GmbH 
constructed the pipes and the wind supply system.


\begin{thebibliography}{19}
\providecommand{\natexlab}[1]{#1}
\providecommand{\url}[1]{\texttt{#1}}
\expandafter\ifx\csname urlstyle\endcsname\relax
  \providecommand{\doi}[1]{doi: #1}\else
  \providecommand{\doi}{doi: \begingroup \urlstyle{rm}\Url}\fi

\bibitem[Rayleigh(1882)]{Rayleigh-82}
J.~W.~S. Rayleigh.
\newblock On the pitch of organ-pipes.
\newblock \emph{Phil. Mag.}, XIII:\penalty0 340--347, 1882.

\bibitem[Pikovsky et~al.(2001)Pikovsky, Rosenblum, and
  Kurths]{Pikovsky-Rosenblum-Kurths-01}
A.~Pikovsky, M.~Rosenblum, and J.~Kurths.
\newblock \emph{Synchronization---A Universal Concept in Nonlinear Science}.
\newblock Springer, Berlin, 2001.

\bibitem[Fletcher(1999)]{Fletcher-NL}
N.~H. Fletcher.
\newblock The nonlinear physics of musical instruments.
\newblock \emph{Rep. Prog. Phys.}, 62:\penalty0 723--764, 1999;
N.~H. Fletcher.
\newblock Mode locking in non-linearly excited inharmonic musical oscillators.
\newblock \emph{J. Acoust. Soc. Am.}, 64:\penalty0 1566--69, 1978.

\bibitem[Rossing and Fletcher(1998)]{RossingFletcher-97}
T.~Rossing and N.~Fletcher.
\newblock \emph{The physics of musical instruments}.
\newblock Springer, New York, 1998.

\bibitem[Succi(93)]{CAA}
S.~Succi.
\newblock \emph{The lattice Boltzmann equation for fluid dynamics and beyond}.
\newblock Oxford University Press, New York, US;
J.~C. Hardin and M.~Y. Hussaini.
\newblock \emph{Computational Aeroacoustics}.
\newblock Springer, Berlin, 1993.

\bibitem[Howe(2003)]{Howe-03}
M.~S. Howe.
\newblock \emph{Theory of vortex sound}.
\newblock Cambridge texts in applied mathematics. Cambridge University Press,
  Cambridge, UK, 2003.

\bibitem[Fabre and Hirschberg(2000)]{Fabre-00}
B.~Fabre and A.~Hirschberg.
\newblock Physical modeling of flue instruments: A review of lumped models.
\newblock \emph{Acustica - Acta Acustica}, 86:\penalty0 599--610, 2000.

\bibitem[Nayfeh and Mook(1979)]{Nayfeh-Mook-79}
A.~Nayfeh and D.~Mook.
\newblock \emph{Nonlinear Oscillations}.
\newblock John Wiley, New York, 1979.

\bibitem[Fabre et~al.(1996)Fabre, Hirschberg, and
  Wijnands]{Fabre-Hirschberg-96}
B.~Fabre, A.~Hirschberg, and A.~P.~J. Wijnands.
\newblock Vortex shedding in steady oscillation of a flue organ pipe.
\newblock \emph{Acustica - Acta Acustica}, 82:\penalty0 863--877, 1996.

\bibitem[Thwaites and Fletcher(1983)]{Thwaites-Fletcher-83}
S.~Thwaites and N.H. Fletcher.
\newblock Acoustic admittance of organ pipe jets.
\newblock \emph{J. Acoust. Soc. Am.}, 74:\penalty0 400--408, 1983.

\bibitem[Arnold(1991)]{Arnold-91}
V.~I. Arnold.
\newblock Cardiac arrhythmias and circle mappings[sup a)].
\newblock \emph{Chaos: An Interdisciplinary Journal of Nonlinear Science},
  1\penalty0 (1):\penalty0 20--24, 1991.

\bibitem[Sauer et~al.(1991)Sauer, Yorke, and Casdagli]{Sauer-Yorke-Casdagli-91}
T.~Sauer, J.~Yorke, and M.~Casdagli.
\newblock Embeddology.
\newblock \emph{J. Stat. Phys.}, 65\penalty0 (3/4):\penalty0 579--615, 1991.

\bibitem[Sch(2004)]{Schuke}
{A}lexander {S}chuke {G}mb{H}, 2004.
\newblock URL \url{http://www.schuke.com}.

\bibitem[Bohn(1988)]{Bohn-88}
D.~A. Bohn.
\newblock Environmental effects on the speed of sound.
\newblock \emph{J. Audio Eng. Soc.}, 36\penalty0 (4):\penalty0 223--231, 1988.

\bibitem[Abel et~al.(2006)Abel, Bergweiler, and
  Gerhard-Multhaupt]{Abel-Bergweiler-Multhaupt-06}
M.~Abel, S.~Bergweiler, and R.~Gerhard-Multhaupt.
\newblock Synchronization of organ pipes by means of air flow coupling:
  experimental observations and modeling.
\newblock \emph{J. Acoust. Soc. Am.}, 119\penalty0 (4):\penalty0 2467 -- 2475,
  2006.

\bibitem[Tufillaro et~al.(1995)Tufillaro, Wyckoff, Brown, Schreiber, and
  Molteno]{Tufillaro-Wyckoff-Brown-Schreiber-Molteno-95}
N.~B. Tufillaro, P.~Wyckoff, R.~Brown, T.~Schreiber, and T.~Molteno.
\newblock Topological time series analysis of a string experiment and its
  synchronized model.
\newblock \emph{Phys. Rev. E}, 51\penalty0 (1):\penalty0 164--174, 1995.

\bibitem[Ahnert and Abel(2007)]{Ahnert-Abel-07}
K.~Ahnert and M.~Abel.
\newblock Numerical differentiation: global versus local methods.
\newblock \emph{Comput. Phys. Commun.}, 177\penalty0 (10):\penalty0 764--774,
  2007.

\bibitem[Abel et~al.(2005)Abel, Ahnert, Mandelj, and
  Kurths]{Abel-Ahnert-Mandelj-Kurths-05}
M.~Abel, K.~Ahnert, S.~Mandelj, and J.~Kurths.
\newblock Additive nonparametric reconstruction of dynamical systems from time
  series.
\newblock \emph{Phys. Rev. E}, 71:\penalty0 15203, 2005.

\bibitem[Gouesbet and Letellier(1994)]{Gouesbet-Letellier-94}
G.~Gouesbet and C.~Letellier.
\newblock Global vector-field reconstruction by using a multivariate polynomial
  ${L}_2$ approximation on nets.
\newblock \emph{Physical Review E}, 49\penalty0 (6):\penalty0 4955, 1994.

\bibitem[Kralemann et~al.(2008)Kralemann, Cimponeriu, Rosenblum, Pikovsky, ,
  and Mrowka]{Kralemann-Cimponeriu-Rosenblum-Pikovsky-Mrowka-08}
B.~Kralemann, L.~Cimponeriu, M.G. Rosenblum, A.S. Pikovsky, , and R.~Mrowka.
\newblock Phase dynamics of coupled oscillators reconstructed from data.
\newblock \emph{Phys. Rev. E}, 77:\penalty0 66205, 2008.

\bibitem[Cox and D'{A}ntonio(2004)]{Cox-DAntonio-04}
T.~Cox and P.~D'{A}ntonio.
\newblock \emph{Acoustic Diffusers and Absorbers: Theory, Design and
  Application}.
\newblock Taylor \& Francis, London, 2004.

\end{thebibliography}

\end{document}